# Evidences of Sub-Nanometer Orbital Diffusion Lengths in Heavy Metals using Terahertz Emission Spectroscopy

Tongyang Guan[1,2,§], Jiahao Liu[1,2,§], Wentao Qin[1,2], Yongwei Cui[1,2], Shunjia Wang[1,2], Yizheng Wu[1,2,*], Zhensheng Tao[1,2,*]

[1]*State Key Laboratory of Surface Physics and Key Laboratory of Micro and Nano Photonic Structures (MOE), Department of Physics, Fudan University, Shanghai 200433, P R China.*

[2]*Shanghai Key Laboratory of Metasurfaces for Light Manipulation, Fudan University, Shanghai 200433, P R China.*

§ These authors contributed equally to this work.

* Corresponding authors: wuyizheng@fudan.edu.cn; zhenshengtao@fudan.edu.cn.

**Abstract**

The orbital angular momentum of electrons offers a promising, yet underexplored, degree of freedom for ultrafast, energy-efficient information processing. As the foundation of orbitronics, understanding how orbital polarizations propagate and convert into charge currents is essential but remains elusive due to the challenge in disentangling orbital and spin dynamics in thin films. While some theoretical studies predict that orbital transport is constrained to sub-atomic-layer scales in materials, recent experiments have reported exceptionally long orbital diffusion lengths. To address this contradiction, we combine terahertz emission spectroscopy with a wedge-sample platform to systematically investigate spin and orbital transport in heavy metals with sub-nanometer resolution. Our measurements access the previously unexplored thin-film regimes (<3 nm), uncovering anomalous behaviors that challenge the prevailing interpretations of long-range orbital transport. We consistently find the orbital diffusion lengths ($\lambda_L$) to be substantially shorter than the spin diffusion lengths ($\lambda_S$) in heavy metals, with $\lambda_L$ in W approaching 0.36 nm. Interface-sensitive control experiments further *rule out* interfacial orbital-to-charge conversion as the dominant mechanism, supporting the bulk inverse orbital Hall effect as the primary conversion process.

Orbitronics is an emerging field that seeks to control and manipulate electronic orbital angular momentum (*L*) in solids, offering new avenues for information processing and magnetic control[1–4]. Compared to spin angular momentum (*S*), which underpins conventional spintronics[5–7], orbital degrees of freedom can carry larger angular momentum. This enables more efficient torques on magnetic moments, a possibility that has been both theoretically predicted[8–10] and experimentally demonstrated[11–15].

Several experiments have reported long-range orbital transport, suggesting that orbital currents – flows of *L* – can propagate over tens of nanometers and drive magnetic excitation far from material interfaces[11,15–17]. In particular, femtosecond laser-induced terahertz emission spectroscopy has revealed orbital diffusion lengths ranging from 9 nm to 80 nm in non-magnetic materials[18–20], and attributed the observed terahertz conversion to interfacial orbital-to-charge processes mediated by the inverse orbital Rashba-Edelstein effect (IOREE) [18–21].

However, these experimental claims stand in contrast to recent theoretical predictions. While long orbital transport has been proposed under specific momentum-space conditions[22], recent first-principles calculations indicate that orbital propagation is constrained by crystal symmetry, limiting orbital diffusion length to sub-atomic-layer scales[23–25]. The origin of this discrepancy remains unresolved and may involve both experimental and theoretical factors. In particular, commonly used electrical[12,16,17,26] or optical[11,15,18–20] probes cannot unambiguously distinguish between spin and orbital contributions. While X-ray[27] and electron[28] magnetic circular dichroism may offer selective sensitivity to $S$ and $L$, direct observation of pure orbital transport remain elusive.

In light of theoretical predictions that orbital transport is limited to very short length scales[23–25], it is essential to examine orbital transport in thin regimes (<3 nm) with sub-nanometer precision, where signatures of orbital diffusion are expected to emerge. In this work, we systematically investigate orbital and spin transport in heavy metals (HMs) using terahertz emission spectroscopy on wedge-shaped HM|ferromagnet (FM) heterostructures. The wedge-sample platform provides sub-nanometer spatial resolution and enables continuous thickness scanning, allowing access to the previously unexplored thickness range of 0-3 nm. Our measurements uncover anomalous behaviors that directly challenge prevailing interpretations of long-range orbital transport.

Specifically, in W|Ni heterostructures, we observe a non-monotonic thickness dependence of terahertz amplitude and a polarity reversal in the thin-W regime, with the orbital diffusion length $\lambda_L$ in W approaching ~0.36 nm, much shorter than the spin diffusion length of $\lambda_S \approx 2.20$ nm. Furthermore, our high-resolution, interface-sensitive control experiments *rule out* the dominant roles of interfacial conversion mechanisms, supporting bulk-origin scenarios mediated by the inverse orbital Hall effect (IOHE) and inverse spin Hall effect (ISHE). A multi-component terahertz emission model is established and enables semi-quantitative extraction of $\lambda_L$ across several HMs (W, Ta, and Pt), consistently showing $\lambda_L < \lambda_S$. These findings provide important experimental constraints on a key controversy in orbitronics and establish a robust experimental

framework for disentangling spin and orbital dynamics at the nanoscale.

## Experimental Principle

Upon excitation by a femtosecond laser pulse, ultrafast spin ($j_S$) and orbital ($j_L$) currents are simultaneously generated in the FM layer of a HM|FM heterostructure with a wedge-shaped HM film (Fig. 1). The $S$ and $L$ polarizations are aligned by an external magnetic field ($H$), and the resulting currents propagate along the out-of-plane ($z$) direction into the adjacent HM layer, where spin-to-charge and orbital-to-charge conversion processes give rise to in-plane charge currents that emit terahertz radiation.

To disentangle spin and orbital contributions, we selected FM and HM materials with distinct current-generation and conversion characteristics. Ni and Fe are selected as FM metals due to their high efficiency in generating orbital and spin currents, respectively[29,30]. For HM layers, we employ W, Ta, and Pt, which possess distinct signs of spin ($\gamma_{\mathrm{SH}}$) and orbital ($\gamma_{\mathrm{OH}}$) Hall angles[9,31,32], as well as differing orbital and spin diffusion lengths ($\lambda_L$ and $\lambda_S$).

In this study, we fabricated wedge-shaped heterostructures where the HM thickness ($d_{\mathrm{HM}}$, HM=W, Ta, Pt) varies continuously along the in-plane ($x$) direction, with a thickness gradient of ~1.6 nm/mm (Fig. 1). The FM-layer thickness ($d_{\mathrm{FM}}$, FM=Ni, Fe) is fixed at 5 nm. By translating the sample in 0.1 mm steps along the $x$ direction, we scan $d_{\mathrm{HM}}$ with a resolution of ~0.16 nm, defined by the laser spot radius of ~50 μm. Terahertz signals are detected via free-space electro-optic sampling. Beyond enabling continuous thickness sampling, the wedge-sample platform mitigates ambiguities associated with growth-condition variations between different samples. See Methods and Supplementary Section S1 and S2 for further details and comparison with thin-film samples of constant $d_{HM}$.

**W|Ni and Ni|W heterostructures**

Figure 2 presents terahertz emission measurements from wedged-shaped W|Ni and Ni|W heterostructures. For comparison, we also include their Fe-based counterparts, W|Fe and Fe|W. Throughout this manuscript, "HM|FM" denotes the growth sequence of the heterostructure on the substrate, with the HM layer deposited first, followed by the FM layer. For clarity, the $SiO_2$ buffer and capping layers are omitted from the notation unless otherwise specified. The time origin ($t$=0) is consistently defined by the main peak of the W|Fe or Fe|W reference sample.

Figure 2a show selected terahertz waveforms from W(5)|Ni, W(0.8)|Ni, and W(5)|Fe (thickness in parentheses are in nanometers). The corresponding Ni|W and Fe|W signals are shown in Fig. 2b. In W|Fe, the terahertz polarity is consistent with the spin Hall angle ($\gamma_{SH}$) of W and opposite to that of Pt|Fe, indicating that terahertz generation is predominantly driven by spin currents via ISHE[33–35]. In Fe|W, the polarity is reversed due to the opposite stacking order, as expected.

In contrast to the Fe-based heterostructures, the Ni-based samples (W|Ni and Ni|W) exhibit different behaviors with varying W-layer thickness.

For W|Ni, the terahertz polarity at $d_W \approx 0.8$ nm is opposite to that of W|Fe, indicating that ISHE cannot be the dominant emission mechanism in the thin-W regime. As $d_W$ increases, the W|Ni signal undergoes a polarity reversal, such that the polarity of W(5)|Ni aligns with that of W(5)|Fe (Fig. 2a). This crossover suggests a transition from

a non-ISHE mechanism at small $d_W$ to the ISHE in the thick-W regime.

For Ni|W, a similar polarity reversal is observed: the Ni|W(0.8) sample exhibits polarity opposite to Fe|W(0.8), while for $d_W \approx 5$ nm, the polarity at $t$=0 becomes aligned (Fig. 2b). However, different from W|Ni, this reversal is accompanied by a pronounced waveform change corresponding to a phase shift that effectively enhances the negative peak near $t \approx 0.3$ ps, consistent with earlier observations[18] (Fig. 2b).

Figures 2e-h present the full evolution of terahertz waveforms as a function of $d_W$, with the corresponding amplitudes extracted at $t$=0 plotted in Figs. 2c-d. For the Fe-based samples, the amplitude increases with $d_W$ and peaks around $d_W \approx 3$ nm, consistent with the characteristic behavior of spintronic emitters[35,36]. Fitting the amplitude trend with the diffusive transport model[35] yields a spin diffusion length in W of $\lambda_S \approx 2.20$ nm.

For W|Ni and Ni|W heterostructures, our continuous wedge-sample measurements uncover four key common features previously unresolved. First, for W thickness below 2.8 nm, both W|Ni and Ni|W exhibit terahertz polarities at $t$=0 opposite to those of the corresponding Fe-based counterparts, indicating that a dominant conversion mechanism *distinct from* ISHE in this regime. Second, both stacking orders exhibit identical intensity features: a peak at ~0.8 nm and a minimum around 2-3 nm (Fig. 2i). Third, as the W thickness increases beyond ~2.8 nm, the terahertz polarity at $t$=0 reverses and becomes aligned with that of the Fe-based counterparts, suggesting a crossover to ISHE-dominated emission. Fourth, the associated peak-time delays are confined to a narrow thickness window around the emission-intensity minimum at $d_W \approx 2$-3 nm (see Fig. 2e, g and i, and Supplementary Fig. S11).

Despite these shared behaviors, stacking-order-dependent differences in waveform shape are observed. In Ni|W, the polarity reversal at $d_W$≈2.8 nm is accompanied by an additional phase offset relative to W|Ni, such that the signals from both stacking orders are not strictly opposite to each other (Figs. 2a-b). While the origin of this offset remains unclear, it does not affect the common thickness-dependent trends and therefore does not alter the identification of the dominant emission mechanisms. By contrast, W|Ni exhibits a polarity reversal with minimal phase distortion (Fig. 2e), providing a more direct view of the intrinsic crossover between orbital- and spin-mediated terahertz emission in W.

**Exclusion of Interfacial Mechanisms**

The non-monotonic variation of terahertz amplitude for $d_W$<3 nm in both W|Ni and Ni|W indicates that neither pure spin nor orbital transport alone can account for the data. The sharp amplitude increase below $d_W$≈0.8 nm suggests the involvement of interfacial Rashba-type conversion mechanisms, such as IOREE[18–21] or the inverse spin Rashba-Edelstein effect (ISREE)[37,38].

To assess these interfacial mechanisms, we perform a series of control experiments by inserting a 1-nm Cu-spacer layer at either the W|Ni or $SiO_2$|W interface in the wedge-shaped W|Ni heterostructures (Fig. 3a). This approach intentionally disrupts potential interfacial channels, including the ISREE and IOREE.

In Fig. 3b, we compare the terahertz signals from three configurations: (i) the reference sample $SiO_2$|W(0.7)|Ni, (ii) $SiO_2$|W(0.7)|Cu(1)|Ni with 1-nm Cu inserted at the W|Ni interface, and (iii) $SiO_2$|Cu(1)|W(0.7)|Ni with 1-nm Cu inserted at the $SiO_2$|W

interface. Importantly, both waveform and amplitude remain essentially unchanged upon Cu insertion at either interface. In Fig. 3c, we further present the terahertz amplitudes as a function of $d_W$ for all the three configurations. Notably, all the samples exhibit a consistent trend: a positive terahertz signal for $d_W$<3 nm, peaking around $d_W$≈0.7-0.8 nm, followed by an amplitude decrease with increasing $d_W$. The persistence of both signal strength and its $d_W$-dependence across all configurations essentially *rule out* the IOREE and ISREE as the dominant conversion mechanisms.

Although the Cu-layer insertion inevitably introduces new interfaces, i.e. $SiO_2$|Cu, Cu|W, W|Cu, and Cu|Ni, that could contribute to the signal, several observations exclude this possibility. First, the consistent amplitude increase observed for $d_W$<0.8 nm across all configurations rules out the $SiO_2$|Cu and Cu|Ni interfaces, which are decoupled from W. If either dominated the emission, increasing $d_W$ would suppress the signal due to orbital-current blocking or electrical shunting, contrary to the observations. Second, if IOREE or ISREE at the Cu|W or W|Cu interfaces were the dominant sources, reversing the interface orientation (i.e. $SiO_2$|Cu|W|Ni *vs.* $SiO_2$|W|Cu|Ni) would invert the terahertz polarity. Yet, both configurations exhibit identical polarity and amplitude trends (Fig. 3b), further excluding these interfacial processes.

Together, these findings firmly *exclude* interfacial mechanisms behind the observed terahertz emission. This conclusion is reinforced by additional control experiments incorporating 2-nm Cu and 1-nm Al spacers. In both cases, the characteristic amplitude peak for $d_W$ <3 nm persists, indicating that the key emission features are robust against different spacer materials and interfacial configurations (see

Supplementary Section S4).

Experimentally, we observe peak-time offsets as large as ~250 fs in W|Ni at $d_W$≈4.0 nm (Supplementary Figs. S9a), far exceeding those expected for orbital transport given reported orbital velocity of $v_L$≈0.1-0.26 nm/fs[18,20]. Notably, these offsets appear only within a narrow thickness window of 2.0-3.5 nm, coinciding with the minimum in terahertz amplitude (Figs. 2e, g, and i). In contrast, such time offsets are absent in Ta|Ni or Pt|Ni, where no amplitude minimum is observed (see below). These observations are inconsistent with an interpretation based on delayed interfacial conversion due to orbital transport[18], and instead support a scenario in which the peak-time offsets arise from destructive interference between spin and orbital contributions of opposite polarity. This interpretation is consistent with the opposite signs of the spin and orbital Hall angles in W ($\gamma_{SH}<0$, $\gamma_{OH}>0$)[9,31,32]. Further analysis is provided in Supplementary Sections S5 and S11.

**Disentangling Spin and Orbital Contributions**

Building on the bulk-conversion picture, we further examine the polarity trends. At large W thicknesses ($d_W$>3 nm), the terahertz polarity of W|Ni matches that of W|Fe (Fig. 2a), suggesting that spin-to-charge conversion via ISHE dominates in the thick-W regime. Conversely, at thinner W layers ($d_W$<3 nm), the polarity reverses, suggesting that the IOHE becomes dominant. Consequently, the crossover behavior at $d_W$≈2.8 nm indicates that the orbital diffusion length ($\lambda_L$) is shorter than the spin diffusion length ($\lambda_S$) in W.

To enable quantitative comparisons, we develop a multi-component terahertz emission model (MC-TEM) by extending the established spin diffusion model[35,36,39] to describe orbital propagation. In the absence of a rigorous microscopic theory for orbital transport, we adopt a pragmatic approach by modeling orbital propagation using the same diffusive formalism as spin currents. This approach enables a meaningful, semi-quantitative comparison between spin and orbital diffusion lengths. Within this context, the MC-TEM yields an effective orbital diffusion length ($\lambda_L^{\mathrm{eff}}$), without resolving additional complex processes, such as orbital-to-spin conversion[25,40].

Details of the MC-TEM are provided in Methods. Briefly, the total terahertz emission is modeled as a superposition of three components:

$$E(t, d_{\mathrm{HM}}) = A_S(d_{\mathrm{HM}})\mathcal{E}_S(t) + A_L(d_{\mathrm{HM}})\mathcal{E}_L(t) + A_{\mathrm{Ni}}(d_{\mathrm{HM}})\mathcal{E}_{\mathrm{Ni}}(t), \qquad (1)$$

where the first two terms represent the spin and orbital contributions in the HM, and the third term accounts for emission from the ferromagnetic Ni layer. Experimentally, terahertz emission can be observed even in the absence of the W layer, indicating that Ni layer itself contributes to the emission. This Ni-originated signal can be attributed to the anomalous Hall effect (AHE)[20,41] and magnetic dipole emission (MDE)[42,43], and must be included for a realistic model (see Supplementary Section S3). In the model, each term is decomposed into a thickness-dependent amplitude $A_m$ and a thickness-independent waveform $\mathcal{E}_m$ (m=$S$, $L$, and Ni), supported by our experimental observations (see Methods).

The spin ($A_S$) and orbital ($A_L$) amplitudes are described using diffusive transport equations that incorporate optical absorption, interfacial current reflections, and electrical shunting effects[35]. Because the Ni thickness is fixed in all samples, the Ni-layer contribution ($A_{Ni}$) is treated as a constant current source whose contribution decreases with increasing $d_W$ due to enhanced optical attenuation and electrical shunting. The spin- and Ni-related waveforms, $\mathcal{E}_S(t)$ and $\mathcal{E}_{Ni}(t)$, are directly extracted from the experimental data, while the orbital waveform $\mathcal{E}_L(t)$ is modeled by applying a carrier-envelope phase (CEP) shift $\Delta\varphi$ to $\mathcal{E}_S(t)$, with $\Delta\varphi$ assumed to be thickness-independent (see Methods).

Most variables in the MC-TEM are determined either by reference databases or by direct measurements (see Methods and Supplementary Section S7). In particular, by setting $A_L$=0 and $A_{Ni}$=0, the MC-TEM reduces to the established framework for spintronic emitters, which enables the extraction of spin diffusion lengths $\lambda_S$ from the Fe-based heterostructures (Figs. 2c and d). Consequently, only 4 independent fitting parameters remain for each HM|Ni sample: the effective orbital diffusion length in HM ($\lambda_L^{\mathrm{eff}}$), the spin and orbital amplitude coefficients ($\tilde{A}_S$ and $\tilde{A}_L$), and the CEP offset ($\Delta\varphi$). Notably, $\tilde{A}_S$ and $\tilde{A}_L$ are proportional to the spin and orbital Hall angles ($\gamma_{SH}$ and $\gamma_{OH}$), respectively.

Taking W|Ni as an example, the MC-TEM provides excellent fits across the full measured thickness range ($d_W$= 0 - 15 nm), capturing both amplitude and waveform evolution (Figs. 4a-f and Supplementary Section S8). In particular, the model accurately reproduces the non-monotonic intensity trend (Fig. 4g). Figure 4h presents the extracted amplitudes $A_L$, $A_S$, and $A_{Ni}$ as a function of $d_W$. The fitting results confirm that the intensity minimum arises from destructive interference between the spin and orbital contributions when $A_S + A_L \approx 0$.

In the bulk-conversion framework, diffusion lengths determine the initial rise of the terahertz amplitude at small $d_W$, with a steeper increase corresponding to a shorter diffusion length through the $\tanh(d_W/2\lambda_m)$ dependence[35], where $\lambda_m$ denotes the spin ($\lambda_S$) or effective orbital ($\lambda_L^{\mathrm{eff}}$) diffusion length (Fig. 4h). At larger $d_W$, the amplitude decrease is governed by optical absorption and electrical shunting. For W|Ni, the extracted effective orbital diffusion length is $\lambda_L^{\mathrm{eff}} \approx 0.36$ nm – on the order of a single unit cell – much shorter than the spin diffusion length ($\lambda_S \approx 2.20$ nm). The MC-TEM also qualitatively reproduces the anomalous waveform evolution observed in Ni|W, yielding a consistently short $\lambda_L^{\mathrm{eff}} \approx 0.35$ nm (Supplementary Section S10).

We extend this analysis to the Ta|Ni and Pt|Ni heterostructures and their respective Fe-based counterparts (Figs. 5a-f). In Ta|Ni, the terahertz signal exhibits a polarity opposite to that in Ta|Fe (Figs. 5c-d), but aligns with the polarity of Pt|Fe (Fig. 5f). This behavior is inconsistent with the ISHE given that Ta and Pt have opposite spin Hall angles[9,31,32], suggesting the dominant orbital contribution in Ta|Ni[20]. Moreover, the sharper amplitude rise and the earlier peak in Ta|Ni compared to Ta|Fe (Fig. 5a) indicate a shorter orbital diffusion length. In contrast, Pt|Ni exhibits no polarity reversal (Figs. 5e-f), consistent with the same sign of spin and orbital Hall angles in Pt[9,31,32]. Nonetheless, the amplitude maximum occurs at a smaller thickness than in Pt|Fe (Fig. 5b), again indicating a shorter characteristic diffusion length when orbital contributions are involved.

The MC-TEM reproduces both the amplitude and waveform data across all systems (see Supplementary Section S8), with the extracted parameters summarized in

Table 1. To account for potential amplitude variations between measurements, the spin and orbital amplitudes ($\tilde{A}_S$ and $\tilde{A}_L$) are normalized to the Ni-layer emission amplitude ($b$). Across all three HM materials (W, Ta, and Pt), the extracted orbital diffusion lengths are consistently shorter than their spin counterparts ($\lambda_L^{\mathrm{eff}} < \lambda_S$). For Ta and Pt, $\lambda_L^{\mathrm{eff}}$ is 0.84 and 0.94 nm, respectively, spanning several unit cells.

Figure 5g further summarizes the extracted spin and orbital amplitudes $(\tilde{A}_S/b)$ and $(\tilde{A}_L/b)$ across all three HM|Ni systems. The signs of the amplitudes are consistent with the first-principles predictions for those of the intrinsic spin and orbital Hall conductivities ($\sigma_{\mathrm{SH}}$ and $\sigma_{\mathrm{OH}}$) of each HM (Fig. 5h)[9,31,32], providing further validation of our model.

While recent theoretical studies propose that extrinsic mechanisms can contribute to the orbital Hall effect in disordered or engineered systems[44–46], our results show clear agreement on the relative signs of the *intrinsic* $\sigma_{\mathrm{SH}}$ and $\sigma_{\mathrm{OH}}$ predicted for bulk heavy metals (Fig. 5g-h). This consistency suggests that, in our case, spin and orbital transport are primarily governed by intrinsic material properties.

We further note that the accurate reproduction of the terahertz waveforms requires the inclusion of the CEP offset $\Delta\varphi$, particularly for W|Ni near $d_{\mathrm{W}}\approx$2-3 nm, where spin and orbital contributions nearly cancel (Figs. 4c-d). In Ni|W, a thickness-dependent $\Delta\varphi$ captures the overall trends but not all waveform details (Supplementary Section S10), suggesting that phase offsets may be influenced by extrinsic factors such as interface quality or growth conditions, which merit further investigation.

## Ruling Out Internal Conversion Processes in Ni

The anomalous behaviors observed in HM|Ni heterostructures for $d_{HM}$<3 nm, particularly the non-monotonic amplitude in W|Ni, raise the question of whether these features could originate from internal spin-to-charge conversion processes within Ni that are modified by proximity to the heavy metal. Such effects have been discussed in the context of self-induced spin-orbit torques[47], spin-current dissipation in adjacent layers[48], and interface-modified anomalous Hall responses[41,49].

These possibilities can be effectively ruled out by spacer-layer-insertion control experiments. As shown in Fig. 3c (and Supplementary Section S4), the insertion of 1-nm Cu, 2-nm Cu, or 1-nm Al at the W|Ni interface does not suppress the terahertz amplitude peak around $d_W \approx 0.8$ nm, nor does it significantly shift its position. Because these spacers strongly disrupt the coupling between the W and Ni layers, the persistence of the peak is incompatible with the interpretation that the observed phenomena arise from Ni-internal processes modulated by W.

Further evidence is provided by measurements on Ta|Ni heterostructures. Ta is a heavy metal with strong-orbit coupling (SOC) comparable to W. If the internal processes in Ni dominate, Ta|Ni should exhibit similar non-monotonic amplitude and polarity reversal as W|Ni. In contrast, Ta|Ni displays a monotonic increase that peaks near $d_{Ta} \approx 2$ nm and decays without polarity inversion (Fig. 5a).

Together, these control experiments demonstrate that the observed anomalous behaviors do not arise from the internal spin-to-charge conversion processes in Ni, but are instead intrinsically associated with the W layer.

**Conclusions**

In this work, we demonstrate that orbital transport in W is confined to sub-nanometer length scales, suggesting that the orbital response is more appropriately described as a localized orbital polarization rather than a propagating orbital current. Within this framework, the extracted effective orbital diffusion length ( $\lambda_L^{\mathrm{eff}}$ ) characterizes the spatial decay of orbital polarization, rather than a microscopic mean free path. This conclusion contrasts with earlier reports of long-range orbital transport[18], and is enabled by resolving a non-monotonic terahertz amplitude in the thin-W regime (0-3 nm) through continuous, sub-nanometer thickness sampling and systematic waveform analysis. Our results rule out interfacial IOREE as the dominant mechanism and demonstrate that long orbital diffusion lengths are incompatible with the observed polarity agreement between W|Ni and W|Fe at large W thicknesses, as well as the confinement of peak time offsets to $d_{\mathrm{W}}\approx 2$ - 3.5 nm. These findings underscore the critical advantage of the wedge-sample approach to resolve subtle thickness-dependent waveform variations and disentangle competing terahertz emission mechanisms. While this study offers new insights into short orbital diffusion lengths in HMs, a complete microscopic understanding remains an open challenge.

First, recent theoretical studies have suggested that orbital-to-spin interconversion can occur in materials with strong SOC, leading to secondary spin currents that contribute to terahertz emission via the ISHE[25,40]. In the present work, such processes are not treated explicitly. Instead, their net effect is captured through an effective orbital diffusion length $\lambda_L^{\mathrm{eff}}$. Given that the extracted $\lambda_L^{\mathrm{eff}}$ is already substantially shorter than $\lambda_S$ in HMs, explicit inclusion of this two-step process would likely further reduce the intrinsic orbital diffusion length. In Ta and Pt, where $\lambda_L^{\mathrm{eff}}$ is comparable to $\lambda_S$, orbital-to-spin conversion may partially contribute to the observed signals. Disentangling this secondary process motivates further studies in materials with weak SOC, such as Ti, Cr, and Cu[11,20,29], or different orbital Hall characteristics.

Second, the spin and orbital diffusion lengths extracted here are shorter than the steady-state spin-flipped diffusion length typically measured in spin pumping experiments[50,51], consistent with previous terahertz emission studies[35,36,39]. This difference likely arise from the hot-electron nature of the laser-induced terahertz emission process, as supported by superdiffusive spin transport simulations[52]. Because spin and orbital transport are energy-dependent, the finding that $\lambda_L<\lambda_S$ in HMs should not be simply generalized to low-frequency or steady-state regimes. Future studies could bridge these regimes by extending terahertz emission spectroscopy toward lower excitation energies and expanding detection bandwidth toward lower frequencies. In parallel, theoretical investigations on the energy dependence of orbital transport would be valuable.

Finally, it is important to revisit and clarify the role of interfacial conversion mechanisms across different experimental contexts. In spin-pumping measurements, orbital signals often originate from the orbital Rashba-Edelstein effect at metal-oxide interfaces; however, recent studies have shown that the reported large orbital Hall conductivities may arise from overlooked spin-related processes[47]. In this work, our experiments clearly rule out IOREE in the terahertz regime, likely because the efficiency of inverse Edelstein effect decreases drastically at terahertz frequencies[53]. These results indicate that disentangling interfacial and bulk contributions across different timescales will be essential for a unified understanding of orbital transport.

**Acknowledgments:** We thank Peter Oppeneer for helpful discussion. This work was accomplished in Fudan University. We acknowledge the support from the National Key Research and Development Program of China (Grant Nos. 2021YFA1400200, 2024FYA1408500 and 2022YFA1404700). Z.T. and Y. W. acknowledge the support from the National Natural Science Foundation of China (Grant No. 12221004), and the Shanghai Municipal Science and Technology (Grant No. 22JC1400200). Z. T. also acknowledges the support from the National Natural Science Foundation of China (Grant Nos. 12450407 and 12274091). Y. W. also acknowledges the support from the National Natural Science Foundation of China (Grant Nos. 12434003 and 12274083).

**Author Contributions Statement:**

Conceptualization: Y. W., Z. T.

Experiment: T. G., J. L., W. Q., Y. C., S. W.

Theory: T. G., J. L., Y. W., Z. T.

Supervision: Y. W., Z. T.

Writing – original draft: T. G., J. L., Z. T., Y. W.

Writing – review & editing: W. Q., Y. C., S. W.

**Competing Interests Statement:** Authors declare that they have no competing interests.

## Tables

**Table 1.** Key parameters extracted using the MC-TEM for different HMs. Error bars represent fitting uncertainties derived from the error landscapes (Supplementary Section S9).

| | W | Ta | Pt |
|---|---|---|---|
| $\lambda_S$ (nm) | 2.20±0.13 | 1.34±0.07 | 2.00±0.34 |
| $\lambda_L^{\mathrm{eff}}$(nm) | 0.36±0.17 | 0.84±0.51 | 0.94±0.44 |
| $\tilde{A}_S/b$ (arb. units) | -50.4±9.0 | -18.2±14.7 | 25.0±20.9 |
| $\tilde{A}_L/b$ (arb. units) | 162.9±25.5 | 107.9±28.4 | 133.4±42.6 |
| $\Delta\varphi$ (rad) | 0.016±0.080 | 0.012±0.147 | -0.045±0.075 |

## Figure Legends/ Captions

**Figure 1. Schematic Illustration of the Experimental Principle.** Femtosecond laser excitations launches $S$ and $L$ propagation from the FM layer into the adjacent HM layer, leading to spin ($j_S$) and orbital ($j_L$) currents. These currents are subsequently converted into transverse charge currents via the ISHE and IOHE, producing terahertz fields $E_S$ and $E_L$, respectively. The HM layer is fabricated with a wedge-shaped thickness gradient along the $x$ direction, enabling continuous variation of HM thickness by scanning the sample position. For illustrative purpose, the figure depicts the transverse charge currents from $j_S$ and $j_L$ as flowing in opposite directions. In reality, their relative directions vary: they are opposite in W and Ta, but are aligned in Pt.

**Figure 2. Terahertz Emission from W|Ni and Ni|W. a.** Representative terahertz waveforms from W(5)|Ni, W(0.8)|Ni, and W(5)|Fe. For clarity, W|Ni waveforms are

vertically offset by +250, and the amplitudes of W(0.8)|Ni and W(5)|Fe are scaled by 30% and 15%, respectively. Signal polarities at $t$=0 for W(5)|Ni and W(5)|Fe are indicated. **b.** Same as **a.**, but for Ni|W and Fe|W heterostructures. The positive peak at $t$=0 and the negative peak at $t$≈0.3 ps of Ni|W(5) are respectively indicated. **c.** Terahertz amplitudes extracted at $t$=0 for W|Ni and W|Fe. Symbols represent experimental data; solid and dashed lines indicate fits based on the MC-TEM. Star symbols denote measurements from discrete thin-film samples. **d.** Same as **c.**, but for Ni|W and Fe|W. **e.-h.** Two-dimensional waveform plots as a function of $d_W$ for W|Ni **(e)**, W|Fe **(f)**, Ni|W **(g),** and Fe|W **(h)**. Color scales indicate terahertz amplitude. **i.** Spectral intensities of W|Ni and Ni|W as a function of $d_W$. Common intensity peaks near $d_W$≈0.8 nm and minimums around 2.8 nm are highlighted.

**Figure 3. Control Experiments Excluding Interfacial Conversion Mechanisms. a.** Schematic illustrations of different configurations. (i) Reference structure $SiO_2$|W|Ni, where the black arrow denotes the spin and orbital currents ($j_S$ and $j_L$), and blue and red arrows indicate ISREE and IOREE currents at the interfaces. (ii) $SiO_2$|W|Cu(1)|Ni, with a 1-nm Cu spacer inserted at the W|Ni interface. (iii) $SiO_2$|Cu(1)|W(0.7)|Ni, with a 1-nm Cu spacer inserted at the $SiO_2$|W interface. **b.** Experimental terahertz waveforms measured from the three sample configurations. **c.** Terahertz amplitudes extracted at $t$=0 as a function of $d_W$ from all configurations. The amplitude peaks near $d_W$≈0.7-0.8 nm are highlighted by a dashed line. **d.** Normalized terahertz waveforms from $SiO_2$|W|Ni heterostructures for different $d_W$, with peak positions indicated by a dashed line.

**Figure 4. Model Fitting of Terahertz Emission Using MC-TEM. a-f.** Representative

fits to the time-domain terahertz waveforms from W|Ni samples with varying W-layer thickness ($d_W$) based on MC-TEM. Symbols represent experimental data; solid lines indicate model fits. **g.** Spectrally integrated terahertz intensity from W($d_W$)|Ni as a function of $d_W$. Experimental data (symbols) are well captured by the model fits (solid line). The grey bar highlights the thickness at which the intensity minimum occurs. **h.** Extracted amplitudes $A_L$, $A_S$, and $A_{Ni}$ as a function of $d_W$. Shaded areas indicate the effective orbital ($\lambda_L^{\text{eff}}$) and spin ($\lambda_S$) diffusion lengths. Signal rising associated with spin- and orbital bulk conversion processes, and the decay due to optical absorption and shunting effects are presented as dashed lines for illustrative purposes. The grey bar marks the W thickness where destructive interference occurs between spin and orbital contributions, i.e., $A_L$+$A_S$≈0, corresponding to the intensity minimum in **g.**

**Figure 5. Distinguishing Spin and Orbital Contributions. a.** Terahertz amplitudes extracted at $t$=0 for Ta|Ni and Ta|Fe. Symbols represent experimental data; solid and dashed lines indicate fits based on the MC-TEM. Peak positions are highlighted by triangle symbols. **b.** Same as **a.**, but for Pt|Ni and Pt|Fe. **c.-f.** Two-dimensional waveform plots as a function of $d_W$ for Ta|Ni **(c)**, Ta|Fe **(d),** Pt|Ni **(e),** and Pt|Fe **(f)**. Color scales indicate terahertz amplitude. **g.** Normalized amplitude coefficients $\tilde{A}_S/b$ (spin) and $\tilde{A}_L/b$ (orbital) as a function of different HM materials, obtained by fitting experimental data using the MC-TEM. Error bars represent fitting uncertainties derived from the error landscapes (Supplementary Section S9). **h.** *Ab-initio* calculated values of the spin ($\sigma_{SH}$) and orbital Hall ($\sigma_{OH}$) conductivities for the corresponding HM materials[9,31,32].

## Methods

**Sample preparation and characterization.** Wedge-shaped HM|FM heterostructures were fabricated on double-side-polished $Al_2O_3$(0001) substrates. Prior to deposition, all substrates were annealed at 150°C for one hour in the sputtering chamber with a base pressure of $2\times10^{-8}$ torr. A 4-nm $SiO_2$ buffer layer was first deposited onto the substrate. The HM|FM heterostructure was then deposit via DC magnetron sputtering at room

temperature under an argon pressure of 3 mTorr. Finally, a 4-nm $SiO_2$ capping layers was deposited to prevent sample oxidation. The pre-deposition of a $SiO_2$ buffer layer ensures that the FM|HM heterostructure is encapsulated between $SiO_2$ layers. $Al_2O_3$(0001) was selected as the substrate material for its high thermal conductivity, which helps mitigate the laser-induced heating and potential sample damage during measurements.

Prior to deposition, the deposition rates of W, Ta, and Pt were systematically calibrated using X-ray reflectometry. To precisely control the HM-layer thickness, wedge-shaped profiles were fabricated by controlling the deposition time during sputtering using a mechanical shutter. The shutter was moved in front of the sample surface at a constant, calibrated velocity, producing a continuous thickness gradient along $x$ in the HM layer, over a lateral distance of 1.0 cm. In this study, the HM-layer thickness gradients were 1.60 nm/mm for W, 2.50 nm/mm for Ta, and 1.33 nm/mm for Pt.

Post-deposition, the spatial variation in the HM-layer thickness along the $x$-axis was validated by measuring the transmission of a near-infrared femtosecond laser beam, confirming the expected thickness gradient (see Supplementary Section S1).

**Experimental setup.** Terahertz radiation was generated using a power-stabilized fiber oscillator (FIBRE SP, Nanjing Keyun Optoelectronics Ltd.), delivering laser pulses with a duration of 110 fs, a repetition rate of 36 MHz, a central wavelength of 1030 nm, and a maximum power of 3.3 W. In our experiments, a typical excitation power of ~0.7 W was used, with the beam incident from the substrate-side of the samples. An external magnetic field ($H$) of 1000 Oe was applied along the $x$-axis to saturate the magnetization of the FM layer.

The emitted terahertz waves were collected and focused by a pair of off-axis parabolic mirrors. The $E_y$ component was selected by a terahertz polarizer, and was subsequently measured using the electro-optic sampling. A 1-mm-thick GaP was used as the detection crystal. To isolate the $H$-dependent signals and exclude potential signals from electrical anisotropy[39], we recorded the differential response by subtracting the signals measured under opposite magnetic-field directions (+$x$ and -$x$).

To investigate the HM-thickness-dependent terahertz emission, the sample was translated laterally along the $x$-axis with a step size of 0.1 mm, enabling spatial mapping across the HM wedge. The excitation beam was focused to a radius of 50 μm. Given the HM-layer thickness gradient of ~1.6 nm/mm, this configuration yields a thickness resolution of ~0.16 nm.

**Multi-component terahertz emission model.** A detailed derivation of the MC-TEM is provided in Supplementary Section S6. By accounting for optical absorption, interfacial reflections of spin and orbital currents, and the electrical shunting from adjacent layers, the terahertz emission amplitudes associated with the spin and orbital currents can be written as[35]:

$$A_{S(L)}(d_{\mathrm{HM}}) = \lambda_{S(L)}^{(\mathrm{eff})} \frac{\gamma_{\mathrm{SH(OH)}} \cdot a_{S(L)}}{d_{\mathrm{HM}} + d_{\mathrm{FM}}} \frac{\tanh\left(d_{\mathrm{HM}}/2\lambda_{S(L)}^{(\mathrm{eff})}\right)}{n_1 + n_2 + Z_0 \cdot (\sigma_{\mathrm{FM}} \cdot d_{\mathrm{FM}} + \sigma_{\mathrm{HM}} \cdot d_{\mathrm{HM}})}, \tag{2}$$

where $a_{S(L)}$ is the injected spin or orbital current density, $n_1$ and $n_2$ are the substrate/air refractive indices, and $\sigma_{\mathrm{FM}}$ and $\sigma_{\mathrm{HM}}$ are the terahertz conductivities of the FM and HM layers, respectively. Here, both the spin and orbital channels are modeled under the same formalism.

For the Ni-layer contribution, since the Ni layer thickness ($d_{Ni}$) is fixed, the transverse terahertz current density in Ni can be treated as a constant current source. After including optical absorption and shunting effects, its amplitude decreases monotonically with increasing $d_{HM}$, and is given by:

$$A_{\mathrm{Ni}}(d_{\mathrm{HM}}) = \frac{b \cdot d_{\mathrm{Ni}}}{d_{\mathrm{HM}} + d_{\mathrm{Ni}}} \frac{1}{n_1 + n_2 + Z_0 \cdot (\sigma_{\mathrm{Ni}} \cdot d_{\mathrm{Ni}} + \sigma_{\mathrm{HM}} \cdot d_{\mathrm{HM}})}, \quad (3)$$

where $b$ is a coefficient proportional to Ni-layer emission amplitude.

The temporal waveforms $\mathcal{E}_m(t)$ are directly determined using experimental measurements. The waveform $\mathcal{E}_{\mathrm{Ni}}(t)$ is directly extracted from the terahertz emission from a 5-nm Ni film and normalized to its peak amplitude (see Supplementary Section S3). For the spin and orbital waveforms, $\mathcal{E}_S(t)$ and $\mathcal{E}_L(t)$, we adopt the experimentally measured waveform from the HM|Fe heterostructures, denoted $\mathcal{E}_S(t)$, as the spin-component waveform. The orbital waveform is modeled by applying a CEP shift $\Delta\varphi$ to the spin waveform:

$$\mathcal{E}_L(t) = \mathcal{E}_S(t) e^{i\Delta\varphi} + c.c. \quad (4)$$

Substituting Eqs. (2-4) into Eq. (1) yields the full MC-TEM, which we use to extract key parameters associated with spin and orbital contributions. Most variables, such as $n_1$, $n_2$, $\sigma_{FM}$, $\sigma_{HM}$, $\lambda_S$, and $b$, can be determined either by reference databases or by direct measurements. Consequently, for each HM|Ni sample, only 4 independent fitting parameters remain: the effective orbital diffusion length in HM ($\lambda_L^{\mathrm{eff}}$), the ISHE/IOHE amplitude coefficients ($\tilde{A}_S = \gamma_{\mathrm{SH}} \cdot a_S$, $\tilde{A}_L = \gamma_{\mathrm{OH}} \cdot a_L$) and the CEP offset ($\Delta\varphi$).

**Parameter determination and fitting procedure.** In the MC-TEM, several key parameters are determined prior to fitting. First, the refractive index of air is fixed at $n_2$=1, while the refractive index of the substrate $n_1$ and the terahertz conductivities $\sigma_{FM}$ and $\sigma_{HM}$ are experimentally determined via terahertz transmission spectroscopy. Detailed experimental results are shown in Supplementary Section S7. Second, the spin diffusion length $\lambda_S$ in HM is independently extracted by fitting the terahertz amplitude from HM|Fe heterostructures as a function of $d_{HM}$, using the MC-TEM by setting $A_L$=0 and $A_{Ni}$=0[35]. The fitting results for W|Fe, Ta|Fe, and Pt|Fe are shown in Figs. 2c, and 5a-b, respectively. Finally, the amplitude coefficient $b$ associated with the AHE+MDE contribution in Ni is determined from the terahertz signal of the HM|Ni wedge-shaped sample evaluated at $d_{HM}$=0, where the signal originates solely from the Ni layer. The above parameter values are provided in Supplementary Table S2.

With these parameters determined, we have 4 independent fitting variables remain for each HM|Ni sample: $\lambda_L^{\mathrm{eff}}$, $\tilde{A}_S$, $\tilde{A}_L$, and $\Delta\varphi$. These parameters are subsequently extracted by performing a global fitting procedure using a genetic algorithm to minimize the deviation between the modeled and experimental terahertz waveforms at different $d_{HM}$.

Notably, the MC-TEM is constrained not only by amplitude *vs.* thickness trends, but also by the full terahertz waveform at each thickness. This high-dimensional data space makes the parameter extraction highly constrained, significantly reducing fitting ambiguity. In Supplementary Section S9, we provide a detailed discussion on the fitting robustness and error landscapes, which yields the error bars in Table 1.

**Data availability:** The minimum dataset is available in the main text, the Supplementary Information and the Supplementary Data. Additional data are available upon request to the corresponding author, Y. W. and Z. T.

**Code availability:** Sources code for multi-component terahertz emission model fitting are provided as the Supplementary Data. Other code used for this study are available upon reasonable request to the corresponding authors, Y. W. and Z. T.